# Laser induced broad band white emission from transparent Cr:YAG ceramics: origin of broadband emission

M. Chaika and W. Strek

*Institute for Low Temperature and Structure Research, Polish Academy of Sciences, Wroclaw, Poland*

**Abstract**

Laser-induced white (light) emission was observed from transparent Cr:YAG ceramics irradiated with a focused continuous wave beam of light from an infrared laser diode. The laser-induced white emission is detected only on the surface of the sample and is not observed in volume. It is found that the intensity of the emission increases exponentially with the laser power density above the threshold. The impact of broadband emission on the power of the transmitted laser beam through the sample was measured. The disappearance of broadband emission due to displacement of the laser beam or an increase in ambient pressure leads to a decrease in the power of transmitted laser beam. Origins of the laser-induced white light emission along with its characteristic features are discussed in terms of multiphoton absorption, intervalence charge transfer and ionic space charge models.

*Keywords:* Cr:YAG transparent ceramics; white light emission; up-conversion; intervalence charge transfer; ionic space charge models

## 1. Introduction

Yttrium aluminum garnet doped with tetravalent chromium ions ($Cr^{4+}$:YAG) are commonly used as a component of tunable solid-state lasers in the 1.35-1.55 μm spectral range or passive Q-switches for laser systems based on YAG doped with rare-earth ions such as $Nd^{3+}$ and $Yb^{3+}$ [1]. However, the lasing efficiency of a high-quality $Cr^{4+}$:YAG crystals is no higher than 10% [2] indicating that another energy losses process of unknown origin is active in these crystals such as Laser-Induced White Emission (LIWE).

Since its discovery in 2009 by Tanner and coworkers, laser induced white emission has been the subject of numerous reports in the last few years [3-11]. This phenomenon has been observed in numerous materials including nanopowder phosphors [4-6,11] nanoceramics phosphors [3,4] graphene [8,12], tungsten [9], bulk crystals [7] etcetera. LIWE was generated under focused laser beam above a threshold in vacuum ambient and is characterized by a high exponent of the power dependence characteristic for multiphoton absorption, long (from a few milliseconds to a few seconds) rise and decay times and relatively low temperature of the light emitting sample. Also, LIWE phenomenon was detected only under vacuum.

In order to explain this phenomenon, the following models have been proposed: black body radiation [13], photon avalanche [5,14], avalanche ionization of tungsten combined with multiplication of electron from ionized tungsten [9], electron-hole recombination [15], $sp^2$-$sp^3$ like hybridization change of graphene foam and graphene nanoparticles [8,12], RE-$O^{2-}$ charge transfer [4], and Inter Valence Charge Transfer [3,6,7,9]. Except black body radiations, all these models were based on multiphoton absorption of the laser light with further transfer of free electrons to the conduction band. These conclusions were based on the fact that an efficient photocurrent accompanies LIWE [3-5,10].

One of the most popular explanations for LIWE phenomenon is based on Intervalence Charge Transfer (INCT) model, especially for Yb-doped phosphors [3,4]. The broadband emission in Cr-doped phosphors, such as the one studied in this article, can be explained in frame of IVCT process. This process has been investigated mostly in mixed valence transition metal compounds and lanthanide ions [16]. IVCT process occurs in mixed valence pairs, such as $Cr^{4+}/Cr^{3+}$. Since mixed valence pairs form a donor–acceptor redox system, electron transfer between the donor and acceptor sites may occur [17]. The multiphoton absorption in $Cr^{4+}/Cr^{3+}$ pair leads to promote an electron transfer between the chromium ions with further broadband visible emission.

Several problems arise when we try to apply this model. First, IVCT can be detected at atmospheric pressure whereas LIWE appears only under vacuum. Secondly, the effect of atmospheric pressure on LIWE is unclear. An increase in ambient pressure can suppress LIWE process as a whole (multiphoton absorption, electron transfer, etc.) or simply suppress radiative emissions. Moreover, IVCT process in chromium ions mixed valence pair leads to appearance of strong absorption centered at 490 nm [18] and no charge transfer luminescence was detected [19]. Based on this, IVCT model should be expanded to explain this mismatch.

This study is aimed at studying LIWE of transparent Cr:YAG ceramics. To the best of our knowledge, here for the first time we report on laser-induced white emission from transparent Cr:YAG ceramics. This allows comparative studies of the properties of LIWE in transparent materials compared to nanopowders and, as result, will improve the model of LIWE.

## 2. Experimental

$Cr^{4+}$:YAG ceramics were obtained from CoorsTek research laboratory located in Uden, Netherland. A photo of the samples is shown in Fig. 1. To carry out the optical investigation, the surface of the transparent Cr:YAG ceramics was polished with diamond abrasive, gradually decreasing the size of the abrasive grains from 30 to 7 μm. After processing, the pellets had a cylindrical shape with a diameter of 15 mm and a thickness of 3.8 mm.

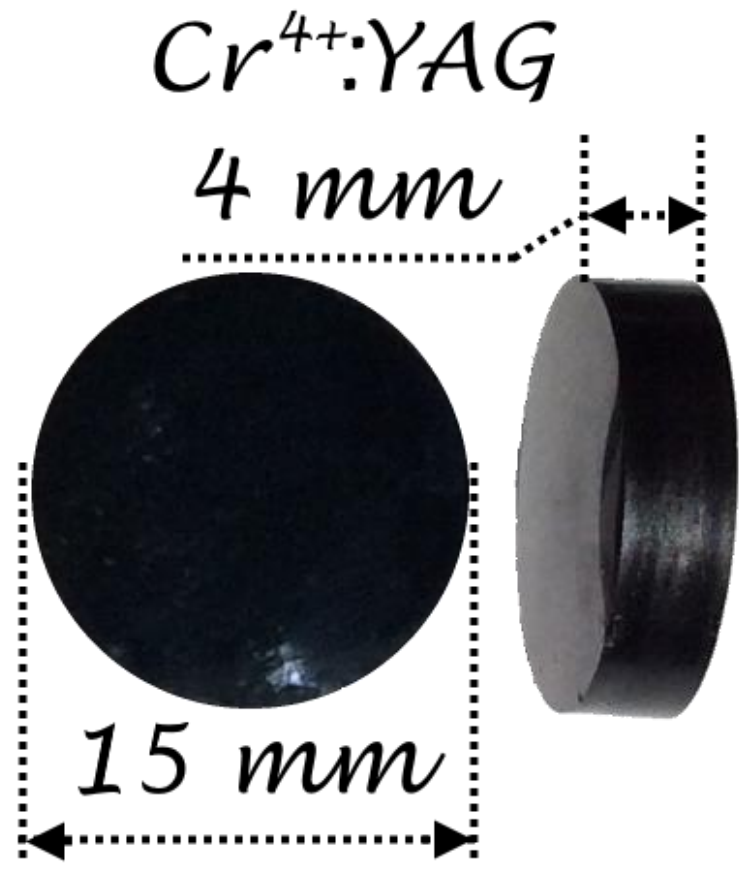

Fig. 1: a) Photo of the mirror-polished $Cr^{4+}$:YAG ceramics.

Optical spectroscopy, and X-ray measurements were used for sample characterizations. The X-ray diffraction spectrum was measured on a Panalytical X'Pert pro X-ray powder diffractometer in the $2\theta$ range of 10°–80° using nickel-filtered Cu $K_{\alpha 1}$ radiation (λ = 1.54056 A) and analyzed by using the WinPlotr Software. The phase composition and cell parameter were calculated by using the Rietveld method.

The measurement of absorbance spectra was performed by a Varian 5E UV-VIS-NIR spectrophotometer with a spectral bandwidth set to 0.5 nm. The emission spectra of the Cr:YAG ceramics were obtained at room temperature by using an Edinburgh Instruments FLS980 fluorescence spectrometer equipped with a 450 W Xe lamp as an excitation source. Two types of detectors were used: a R928P side window photomultiplier tube from Hamamatsu for the visible part of the spectrum and a R5509-72 photomultiplier tube from Hamamatsu in nitrogen-flow cooled housing for near infrared range.

The power of transmitted laser beam through the sample was measured using a USB Power and Energy Meter Interface for C-Type Sensors with Standard Photodiode Power Sensor, Si, 400 - 1100 nm, 500 mW. In order to protect the photodiode power sensor, a 25 x 36mm a shortpass dichroic mirror, with a 900 nm cutoff was used to strongly reduce the power of the transmitted laser beam.

LIWE measurements were performed in vacuum excited by a Nd:YAG laser. The samples were placed in a vacuum cell connected to an EXT75DX Turbo Molecular High Vacuum Pump with an integrated TIC controller (Edwards) in order to perform the measurements at low pressure conditions of $10^{-4}$ mbar. Emission spectra were measured using an infrared continuous wave

Nd:YAG laser as an excitation source. The infrared laser was manufactured by CNI Optoelectronics Tech. Co., Ltd., with central wavelength at 1064 nm, beam divergence <3 mrad, and a RMS power stability of <0.2% . For the collection of emission spectra, two types of detector were used: an AVS-USB2000 Avantes Spectrometer for the visible part of the spectrum and an Ocean Optics NIRQuest512-2.5 Spectrometer for the near infrared range.

The experiment was done in a vacuum chamber (pressure ~$10^{-5}$ mbar) using the 1064 nm excitation beam focused on the sample by a lens to attain a high power density of the laser beam. It is allow us to achieve of cross section of the laser beam up to 0.175 mm. The maximum laser output was 3.4 W. Due to reflections of laser beam at the surface of the lens and vacuum chamber, the power of laser beam at sample surface was reduced by coefficient 0.88. Therefore, the maximum laser output at the surface of the sample was 3.4 W, it is allow reaching the laser density up to $9.8 \cdot 10^3$ W/cm$^2$.

## 3. Result

### *3.1 Structure and optical properties*

Before starting the LIWE experiment, the phase composition of the sample was cheeked. The YAG crystal belongs to the cubic space group Ia3d, the YAG formula can be written as $[C_3][A_2][D_3]O_{12}$, where C, A, and D represent the ligands in a dodecahedral, octahedral, and tetrahedral coordination [20,21]. The A and D sites are occupied by Al ions whereas C sites are taken by Y ions [22,23]. As a single dopant, Cr is incorporated into YAG as a trivalent $Cr^{3+}$ in the octahedral site. In contrast, the tetravalent $Cr^{4+}$ ions can occupy both octahedral and tetrahedral site and requires the presence of divalent impurities such as $Ca^{2+}$ or $Mg^{2+}$ ions in the YAG lattice [20].

Fig. 2a shows the XRD pattern of the Cr:YAG commercial sample. Diffraction data were refined with the cubic Ia3d space group in $Y_3Al_5O_{12}$ by means of the Rietveld analysis. The analysis yields the lattice parameter at 12.0124 ± 0.0005Å. As shown in the XRD pattern and the refinement results, it was evident that all the diffraction peaks can be well indexed as a pure YAG phase, and no secondary phase is observed.

The optical properties of the samples were studied. One of the requirement for transparent Cr:YAG ceramics as material for a Q-switched laser is a high absorption at the laser wavelength and, at the same time, low optical losses that do not occur from $Cr^{4+}$ ions. These requirements defined the properties of commercial transparent Cr:YAG ceramics. The Fig. 2b shown in-line transmission spectra of transparent Cr:YAG ceramics after vacuum sintering and subsequent air annealing. The in-line transmittance of the sample was 3.7 % (8.8 cm$^{-1}$) and 80% (0.16 cm$^{-1}$) at 1064 and 2000 nm respectively. The thickness of the sample was 3.8 mm. A typical transmission spectrum of Cr,Ca:YAG crystal exhibits the broad absorption in the range from the absorption edge to nearly 1100 nm with a minimum of 800 nm. This strong absorption is caused by both the octahedral and tetrahedral $Cr^{4+}$ ions.

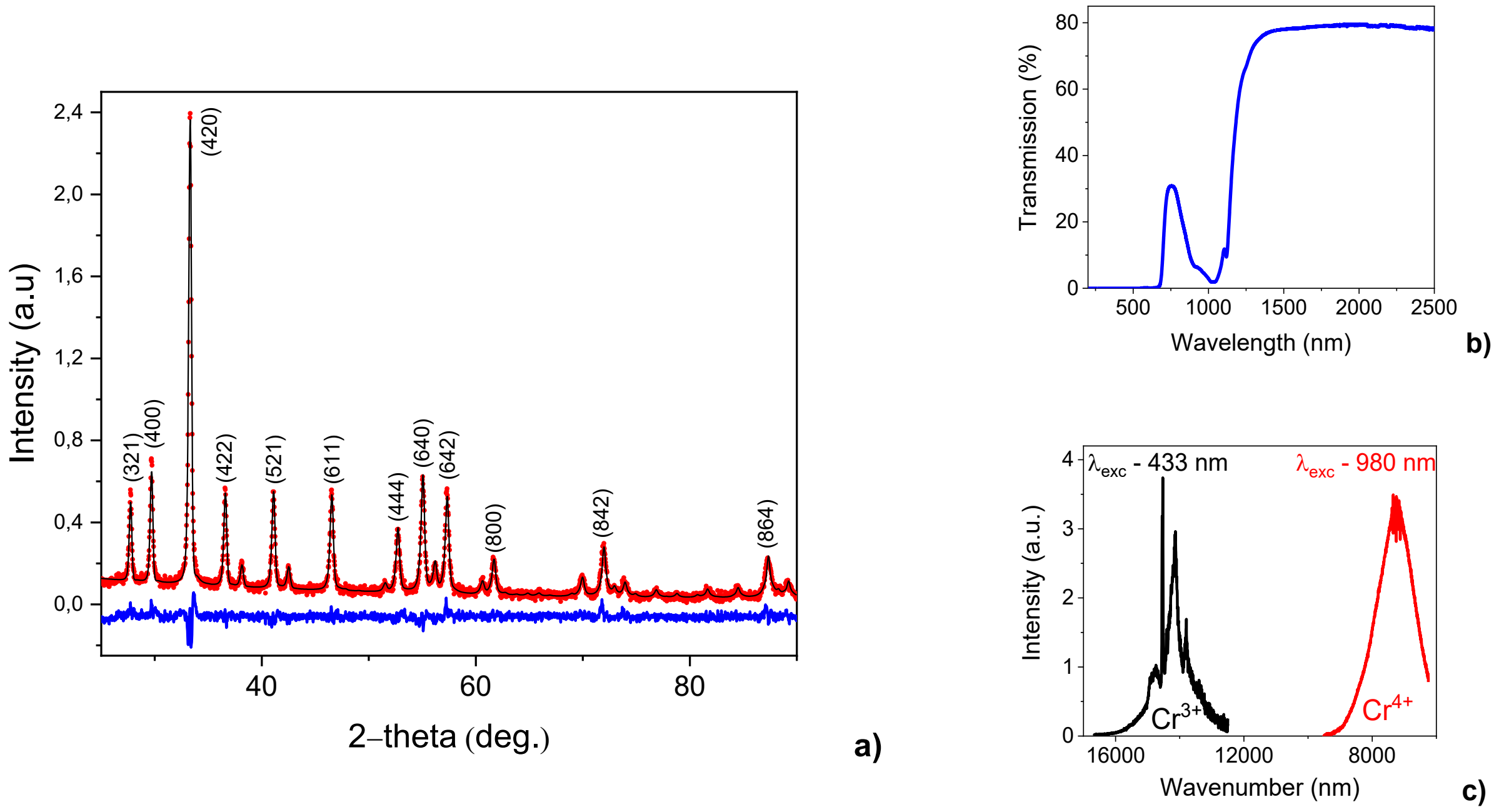

Fig. 2 a) X-ray diffraction pattern of the Cr,Ca:YAG ceramic (black curve), results of Rietveld refinement analysis (red and blue curve); b) transmission spectrum of Cr,YAG ceramic; c) emission spectra of $Cr^{3+}$ (black curve) and $Cr^{4+}$ (red curve) in transparent Cr:YAG ceramic.

The luminescence studies showed the presence of $Cr^{3+}$ and $Cr^{4+}$ ions in the samples. Schematically, the process of making transparent Cr:YAG ceramics can be divided into two stages; vacuum sintering and air annealing. During vacuum sintering, Cr is incorporated into YAG in trivalent state. Annealing in air caused the recharging of only part of $Cr^{3+}$ to the tetravalent state. To prove the presence of $Cr^{3+}$ and $Cr^{4+}$ ions, the luminescent properties of the transparent Cr:YAG ceramics were investigated. Fig. 2c shows the luminescence spectrum of the transparent Cr:YAG ceramics upon excitation with 433 nm (xenon lamp) and 980 nm (diode laser) light. The visible emission comes from the $^2E_g \rightarrow ^4A_{2g}$ and $^4T_{2g} \rightarrow ^4A_{2g}$ transitions of $Cr^{3+}$ ions [24], while the near infrared emission corresponds to the broad band $^3B_2(^3T_2) \rightarrow ^3B_1(^3A_2)$ transition of $Cr^{4+}$ ions in the tetrahedral sites [2].

In order to characterize the samples, the crystal-field strength (Dq/B) was calculated. The crystal-field strength was calculated using the position of $Cr^{3+}$ absorption bands. After vacuum sintering, the ceramic contains chromium only in the trivalent state, as can be seen from the absorption spectrum (Fig. 3a). Annealing the sample in air leads to increased absorption due to formation of $Cr^{4+}$ ions in both octahedral and tetrahedral site (Fig. 3b). According to the Tanabe-Sugano diagram, the samples are characterized by a high crystal-field strength corresponding to Dq/B~2.7

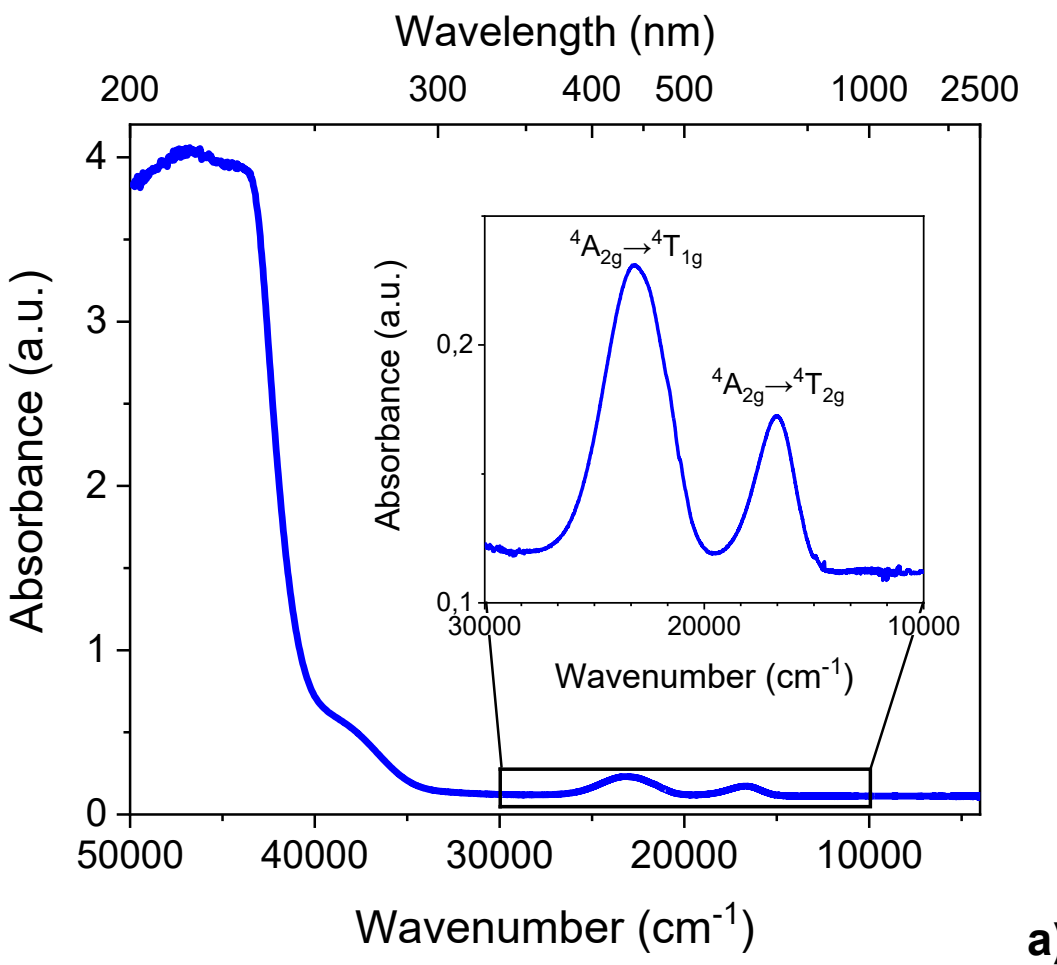

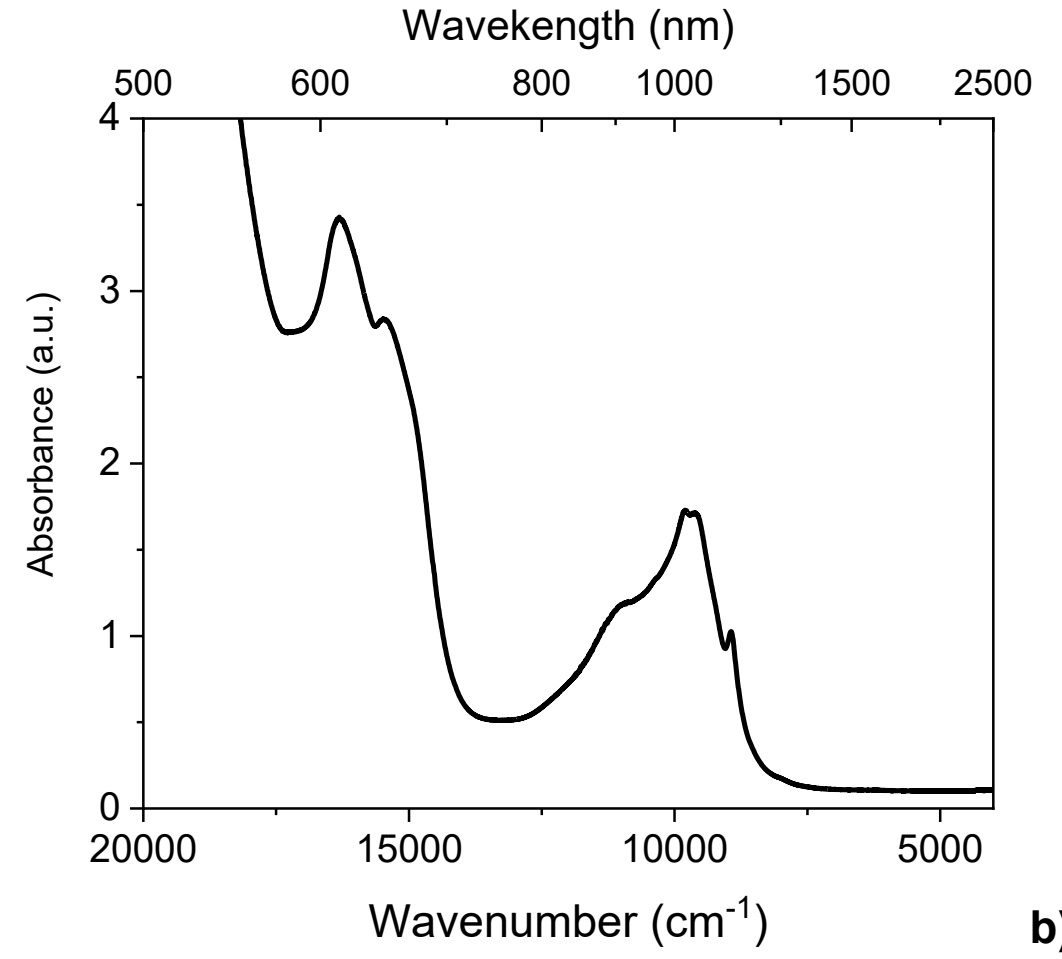

Fig. 3 Absorption spectra of Cr:YAG ceramics after a) vacuum sintering and b) followed air annealing.

### *3.2 Laser Induced White Emission*

The Laser Induced White Emission (LIWE) properties of transparent Cr:YAG ceramics were measured using the 1064 nm laser as the excitation source. The Fig. 4 presents an image of transparent Cr:YAG ceramic the under focused 1064 nm laser light in vacuum ambient. When the excitation light was focused on an appropriate part of the transparent Cr:YAG ceramics with a power above the threshold, the samples were able to generate LIWE. The intensity of LIWE depends on the relative laser spot position on the sample. Some regions couldn't generate LIWE, especially on a smooth lateral surface, where LIWE was observed only a few times. In general, the LIWE intensity from the Cr:YAG ceramics was lower than from nanopowders or nanoceramics [3-6]. According to our knowledge, only one article reports on LIWE from bulk materials [7]. This fact can be explained by the high surface area of nanopowders in comparison with bulk crystals.

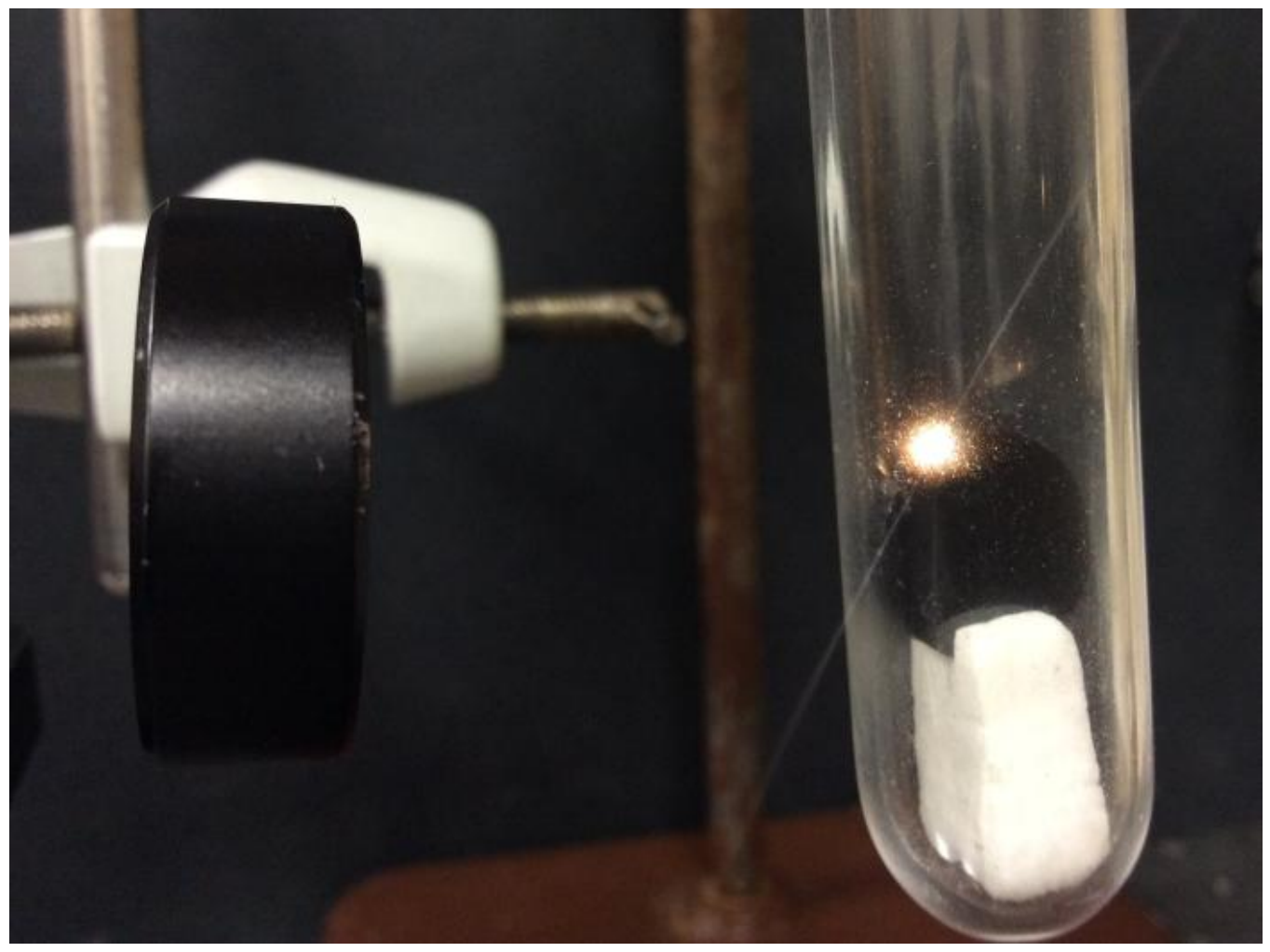

Fig. 4 The photo of LIWE from transparent Cr:YAG ceramics under focused 1064 nm laser excitation in vacuum ambient.

The spectroscopic properties of transparent Cr:YAG ceramics were examined under vacuum conditions (pressure ~$10^{-5}$ mbar) using the 1064 nm excitation beam focused on the sample. Fig. 5 shown the emission spectra for various excitation powers. As can be seen, the visible (VIS) part of the spectrum consists of a broad band emission of maximum localized at 650 nm. Using the same excitation density the near infrared (NIR) part of the spectrum was also recorded, as shown in Fig. 5. The two weak bands located at 1400 and 2200 nm were detected. The first band is attributed to the $Cr^{4+}$ ions, and the second band is of unknown origin.

Due to the fact that VIS and NIR part of emission spectra were measured using two different detectors, the presented spectra were normalized. In addition, the present spectra were uncorrected on the sensitivity of CCD camera. The present spectra demonstrate the possibility of generating visible emission by Cr;YAG ceramics under 1064 nm excitation. The highest LIWE intensity was obtained when the pressure in the measuring chamber was the lowest. An increase in an ambient pressure was accompanied by a decrease of LIWE intensity, which remained stable up to around 0.5 mbar (Fig. 5b). Above this value, the intensity drops sharply and LIWE was not detected at atmospheric pressure.

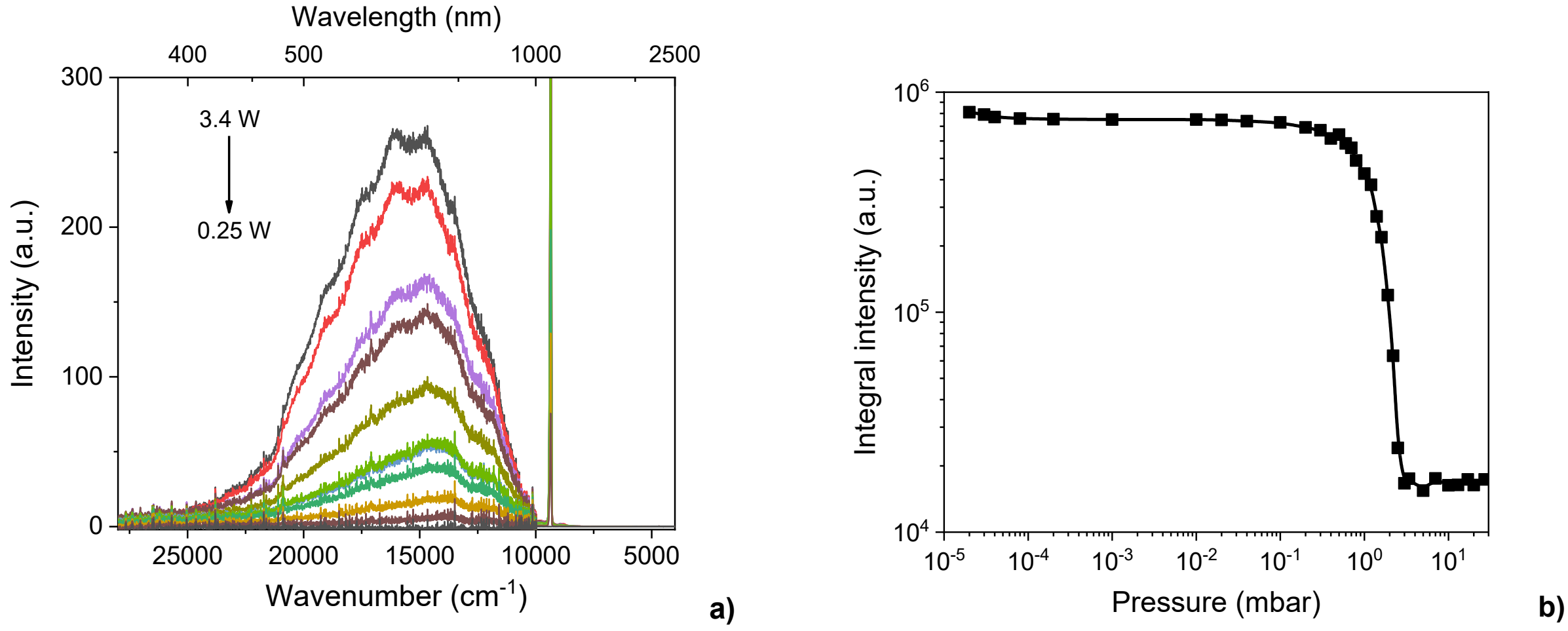

Fig. 5 a) VIS and NIR parts of emission spectra of transparent Cr:YAG ceramics in vacuum under 3.4-0.25 W of 1064 nm excitation. b) The pressure dependence of integral intensity of LIWE observed from transparent Cr:YAG ceramic.

The visible part of the emission spectra highly depends on the excitation density. The power dependence of the emission intensity on the excitation power is shown in Fig. 6 (see also Fig. S2). The integral emission intensity rises by more than two orders of magnitude with increase of the excitation power from 0.1 W to 3.4 W. The LIWE shows the characteristic threshold behaviour. The threshold ranged from 0.3 W to 1.5 W. This dependence may be discussed in terms of a power law expression $I_{em}$~$P^N$ which expresses the intensity (I) as a power function of excited power (P), where exponent (N) denotes the number of photons involved in the emission generation. Above the threshold, a fast increase of LIWE intensity with N in the range from 2.5 to 5.5 for different part of the sample was observed. The white emission in our experiment consists of a single broad band which is located at ~1.9 eV (~15400 $cm^{-1}$). Therefore that emission is characterized by an enormously large Stokes shift in the range from 1 eV to 4.5 eV (from 8000 $cm^{-1}$ to 36300 $cm^{-1}$).

The difference in N for different part of the sample can be explained by the difference in the local concentrations of $Cr^{3+}$ and/or $Cr^{4+}$ ions. The estimated diameter of the irradiation spot was 175 µm. Therefore, it is expected the difference in local distributions of $Cr^{3+}$, and/or $Cr^{4+}$ ions in the transparent Cr:YAG ceramic. It was previously shown that both the N and threshold value depends from concentrations of mixed valence dopants [3-5]. For example, an increase in a concentration of $Yb^{3+}$ ions in Yb:YAG nanocrystals from 2 to 10 at.% leads to increase in N from 4 to 9 and, simultaneously, a threshold reduction from 1.6 W to 1.2 W [3]. High values of N parameter are characteristic of LIWE phenomenon, for example: Yb:YAG nanocrystals (N~4-9) [3], Yb:$Sr_2CeO_4$ nanocrystals (N~4-8), Nd:$Y_2Si_2O_7$ nanocrystals (N~2-6), $LiYbF_4$ nanocrystals (N~6-8) [6].

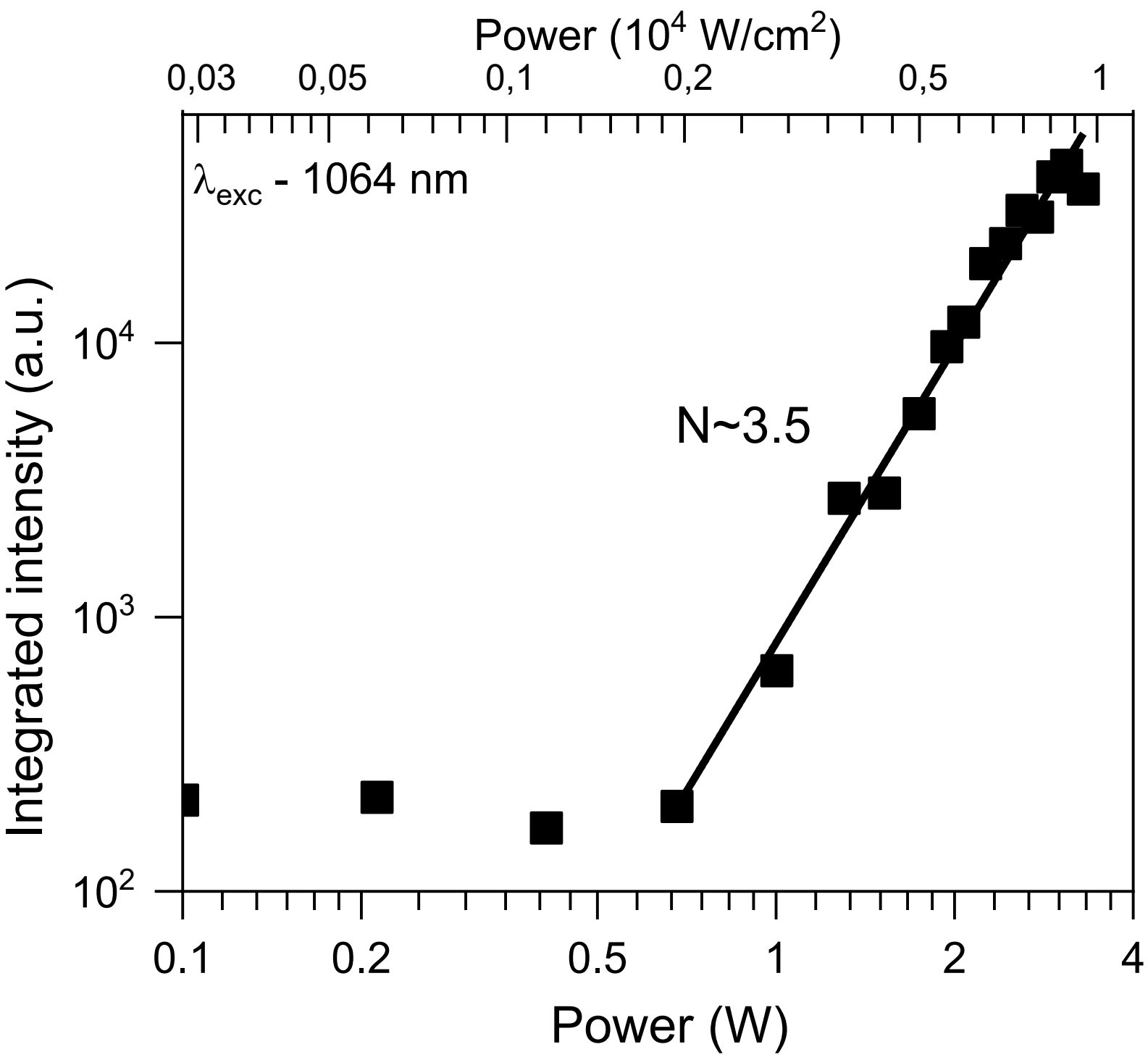

Fig. 6. The influence of the excitation power of the 1064 nm laser diode on the LIWE intensity.

Some authors suggest that the source of LIWE is thermal radiation from the host under the focused laser beam. The energy that collected in the sample enhances the temperature of the host, leading to thermal emission. However, the black body-like emission with intense visible emission should consist of an intense, long emission tail in near infrared part of the spectrum. As can be seen, no evidence of black body emission in NIR is observed in the spectrum of transparent Cr:YAG ceramics under 1064 nm excitation. Moreover the transparent Cr:YAG ceramics have high thermal conductivity, which reduces local accumulation of thermal energy. In addition, damage of sample surface was not observed after 5h of LIWE generation, this indicates relatively low temperature of the sample.

The kinetics of LIWE of transparent Cr:YAG ceramics was measured under vacuum condition and focused CW laser (3.4 W, 1064 nm). The integration time of the CCD camera was 10 ms. Fig. 7a,c show the time evolution of LIWE after switching of the laser. A shift in the emission maximum was not observed in both cases. This result confirms that LIWE did not originate from thermal emission. Fig. 7(b) shows the decay profiles of LIWE in the sample. The experiment was carried out five times in order to exclude the measurements error. This result indicates that the decay time of LIWE is relatively fast and LIWE was absent approximately 20 s after turning on the excitation (Fig. 7b). The decay time was found to be 6±3 ms. In contrary to the decay time profiles, the rise profiles demonstrate some instability after reaching the maximum intensity of LIWE (Fig. 7d). The intensity of LIWE oscillates in some ΔI region. This value of ΔI decreases with time and practically disappears after 0.8s. The rise time was found to be 9±6 ms.

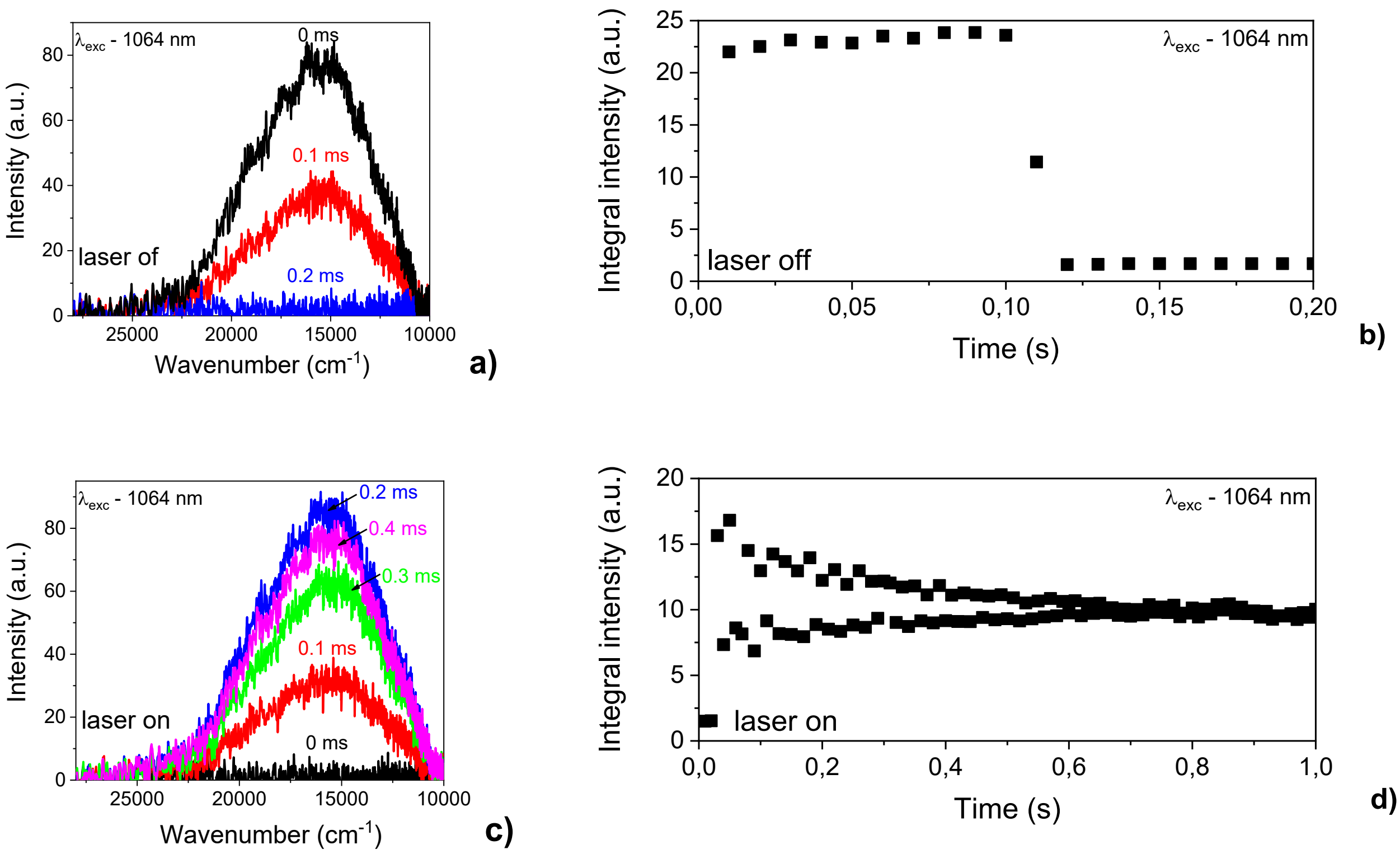

Fig. 7 Temporal evolution of LIWE spectra (a, c) and the integrated intensity of LIWE (b, d) after the switching off (a, b) and switching on (c, d) the excitation laser.

The temporal evolutions in the integrated intensity after switching on and switching off of the incident laser excitation were analyzed using an exponential decay model. The observed decay and rise times were roughly the same. It is important to note that, the detected decay and rise times were much shorter than the earlier reported data. High decay and rise times are characteristic of Laser Induced White Emission (LIWE) process, for example: $LiYbF_4$ nanocrystals (300-600 ms) [6], Yb:YAG bulk crystal (rise time ~ 2900 ms, decay time ~ 170) [7], Yb:$Sr_2CeO_4$ (500-6000 ms) [4], Eu:$Sr_2CeO_4$ nanocrystals (rise time 2000-9000 ms, decay time 40-60 ms) [10]. Approximately the same decay time was found for graphene foam (0.5-0.9 ms) [8] and tungsten filament (4-7 ms) [9].

### *3.3 Impact of LIWE on transmittance*

The main feature of this work in compared to the previous one is the use transparent materials for generation of LIWE. Therefore, the laser beam penetrates the sample during LIWE and exists from opposed side. This made it possible to measure the power of transmitted laser beam through the sample, hereinafter referred to as transmittance. This information give an idea of the effect of LIWE on the absorption of laser beam by Cr:YAG ceramics. Fig. 8 shows the time dependence of the power of transmitted laser beam through the sample during a) displacement of the laser beam or b) increase in ambient pressure. The intensity of the laser is low enough that bleaching of $Cr^{4+}$:YAG is negligible. The measurement scheme is shown in the inset in Fig. 8b.

As it was mentioned before, LIWE was observed only in some places of the sample and at low pressure in the measuring chamber. A shift in the laser beam or an increase in an ambient pressure leads to a change in the LIWE intensity, which causes a change in the transmittance. Therefore, Fig. 8 shows the effect of the presence/absence of LIWE on the transmittance. The Fig. 8a and Fig. 8b were divided into three areas denoted as i), ii), and iii) which correspond to different intensities of LIWE. Region i) was characterised by constant bright LIWE. Region ii) was characterized by a rapid decrease in LIWE intensity due to displacement of the laser beam (Fig. 5a) or an increase an ambient pressure (Fig. 8b). Region iii) did not t show LIWE.

In the case of both methods, a direct effect was observed of the intensity of LIWE on transmittance. Displacement of the laser beam didn't affect the power of transmitted laser beam while LIWE intensity remains the same in both cases - bright LIWE and lack of LIWE (Fig. 8a region i and iii). The disappearance of LIWE leads to a decrease in the power of transmitted laser beam (Fig. 8a region ii)). This difference wasn't the same for different experiments and ranged from 5 to 50%. Similar results were found in the case of increase in ambient pressure in the chamber, with the exception of slightly changing in the transmittance after an increase in an ambient pressure (see Fig. 8a region iii)).

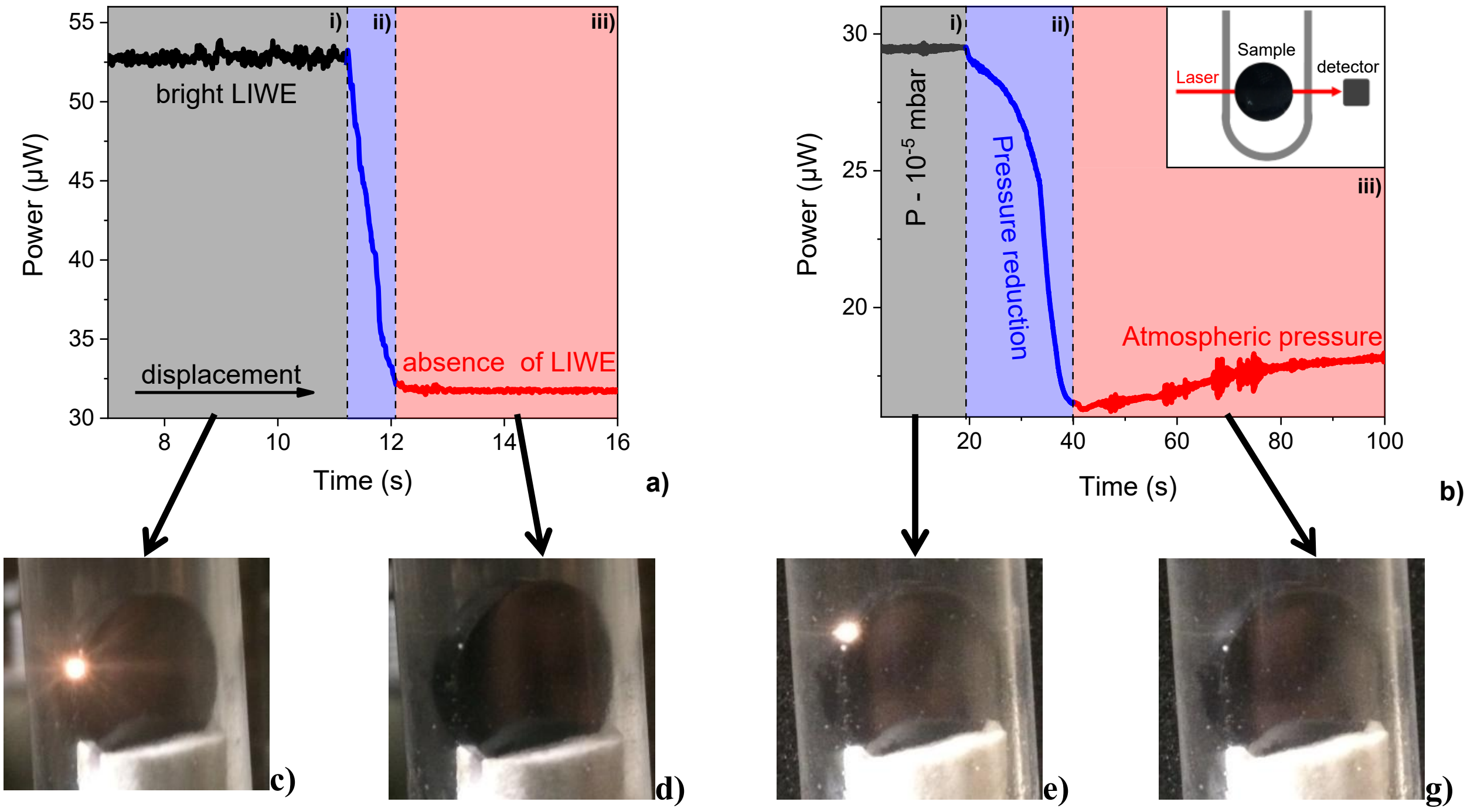

Fig. 8 Time dependence of power of transmitted laser beam through the sample during a) displacement of the laser beam or b) increase in ambient pressure; c-g) photos of the sample during measurements. Inset: the measurement scheme.

In order to understand the role of ambient pressure, the influence of rising pressure on the transmittance of the sample was investigated. Fig. S1 shows the time dependence of the power of the transmitted laser beam through the sample without generating LIWE during an increase of ambient pressure from $10^{-5}$ mbar to $10^3$ mbar. The transmittance remains the same with increasing ambient pressure. Therefore, the decrease in transmittance was caused by disappearance of LIWE and wasn't associated with a change in ambient pressure in measuring chamber.

## 4. Discussion

Laser Induced White Emission (LIWE) has been detected from numerous materials including: graphene foam [8], graphene nanoparticles [12], tungsten filament [9], Yb:YAG nanoceramics [3], Yb:$Sr_2CeO_4$ nanocrystals [4], $Y_2Si_2O_7$:$Nd^{3+}$ nanocrystals [5], $LiYbF_4$ nanocrystals [6], Yb:YAG bulk crystals [7], Cr:YAG and pure YAG nanopowders [11]. However, the mechanism which stands behind the process is still not well understood. The possibility of generating LIWE by different kind of materials indicates that this phenomenon corresponds to some general physical process. Reported results are in close in general (broadband emission, pressure dependence, threshold behavior etc.), but very different in detail. For example, the decay and rise times differ from each other by three orders of magnitude. Also, they have a different value of number of photons involved in the generation of LIWE. But, the emission spectra are close for different materials. Therefore, it is possible that that the emission centers are of the same nature and mechanism of absorption of laser beam is different for different materials.

The several possible mechanisms of LIWE in different materials have been proposed. Most of these mechanisms include excitation of electrons and subsequent recombination by emission of photons. In order to explained the nature of LIWE in transparent Cr:YAG ceramics, the following two stages of this phenomenon should be considered. The first stage is associated with the mechanism that lead to the excitation of electrons, and the second – with the process that is responsible for broadband emission. Proposed mechanisms include: black body radiation [13], photon avalanche [5,14], avalanche ionization of tungsten combined with multiplication of electron from ionized tungsten [9], electron-hole recombination [15], $sp^2$-$sp^3$ like hybridization change of graphene foam and graphene nanoparticles [8,12], RE-$O^{2-}$ charge transfer [4], and Intervalence Charge Transfer [3,6,7,10].

*4.1 Intervalence charge transfer*

Inter Valence Charge Transfer (IVCT) mechanism is the most prevalent explanation for LIWE phenomena in RE/TM-doped oxides. It is not surprising, since most of the results were obtained for Yb-doped oxide, where IVCT process is well studied. IVCT process was previously well described by Seijo et. all [16,17]. Probably the same mechanism is responsible for LIWE in transparent Cr:YAG ceramics and can explain some feature of LIWE phenomenon. Below we will discuss the IVCT process in more detail.

The main feature of LIWE is broadband emission and large Stokes shift. Seijo et al suggested to explain this feature in the frame of IVCT model [16,18]. Since mixed valence pairs form a donor–acceptor redox system, electron transfer between the donor and acceptor sites may occur, and the process can be referred to as IVCT. IVCT processes have been investigated involving mostly mixed valence transition metal compounds and, sometimes, lanthanide ions [17]. A feature of Cr-doped YAG ceramics is the abillity to contain chromium in different valence state, such as $Cr^{2+}$, $Cr^{3+}$, $Cr^{4+}$ and $Cr^{6+}$ [20]. The $Cr^{3+}$ and $Cr^{4+}$ are the most common valence state for such type of ceramics. Therefore, the $Cr^{3+}/Cr^{4+}$ mixed valence pair was considered in this work. The possible combinations of mixed valence pair are not limited to this and may include $Cr^{2+}$, $Cr^{3+}$, $Cr^{4+}$ and $Cr^{6+}$ or even some structural defect.

According to IVCT model, the $Cr^{3+}$ and $Cr^{4+}$ are considered not as separate ions but as the centers of defect including distortions in the first coordination shells due to the difference between ionic radii of Cr dopant and host [17]. Let's denote the two $Cr^{3+}$ and $Cr^{4+}$ ions in the $Cr^{3+}$/ $Cr^{4+}$ ion pair as $Cr_L^{3+}$ and $Cr_R^{4+}$. IVCT in $Cr_L^{3+}/Cr_R^{4+}$ mixed valence pair promoting electron transfer from $Cr_L^{3+}$ to $Cr_R^{4+}$, as a result, $Cr_L^{4+}/Cr_R^{3+}$ mixed valence pair is formed. Caused by difference in the ionic radii of $Cr^{3+}$ (0.062 nm) and $Cr^{4+}$ (0.058 nm) ions, the valence transformation of $Cr^{3+}$ to tetravalent state lead to decrease the Cr-O distance and vice versa. This transformation requires large reorganization energies which is a reason of large Stokes shifts. Also, this indicates that IVCT transition ends in structurally stressed states, which is the reason of observed broad bands emission.

A few problems appear when we look at this model in detail. First of all, two kinds of $Cr^{4+}$ ions exist in YAG lattice. The first is tetrahedral $Cr^{4+}$ ions with strong absorption in the 650-1100 nm region and octahedral $Cr^{4+}$ ions in with strong absorption in the 200-650 nm region [18]. The valence transformation is possible only in the $Cr^{3+}/Cr^{4+}$ mixed valence pair, which both is in the octahedral site [20]. Both octahedral $Cr^{3+}$ and $Cr^{4+}$ ions do not have absorption bands in the near infrared region unlike the $Cr^{4+}$ in tetrahedral site. Moreover, IVCT process between chromium ions leads to strong absorption centered at 490 nm [18] and no charge transfer luminescence has been detected [19]. Therefore, IVCT model must be expanded in order to explain this mismatch.

*4.2 White emission source*

Let us divide LIWE phenomena into two stages. The first stage is associated with the mechanism leading to promotion of electrons into the conduction band, and the second is associated with the process which is responsible for the broadband emission in the visible range. IVCT mechanism may be responsible for multiphoton absorptions of laser light and, as result, promotion of electrons to the conduction band. In order to propose a source of LIWE, it is necessary to outline the role of imperfections (vacancies, impurity ions) in LIWE phenomenon.

Beforehand, LIWE phenomenon has been well studied in the nanopowders. One of the problems of studying LIWE phenomenon is a relatively low purity of the starting source of materials. The nanopowders contain small concentration of uncontrolled impurities originated from the starting source. For example, even high purity YbAG single crystal or Yb;YAG ceramics contain a small fraction of impurities which allow to recharge Yb in divalent state under annealing in a reducing atmosphere [25,26]. These impurities can enable LIWE in nominally undoped sample as well as facilitate broadband emission in RE/TM-doped materials [11]. Also the different kind of imperfections in the crystal structure can also support this phenomenon. But, our result suggested that LIWE was generated only on the surface of transparent Cr:YAG ceramics and no in volume. Therefore, the imperfections of crystal structure weren't a source of broadband emission.

The fact that LIWE was detected on the surface of the Cr;YAG ceramic provides a key to understanding of this phenomenon. The main difference between the volume and surface of oxide is in the charge state. The concentration of oxygen vacancies on the surface of YAG is higher than that in the bulk and lower for metal vacancies [27]. Brown and Bonnell suggested that the YAG surface is charged and this charge are compensated by space charge region (oxygen for example). HRTEM investigation of $Cr^{4+}$:YAG ceramic supports this finding [28]. Several models can be proposed to explain the origin of broadband emission. One of them is electron-hole recombination in this layer with some surface defects, or even oxygen from the space charge region may be responsible for this.

Beforehand, it was shown that oxygen molecules with captured electron trapped in nanoscale cage of $Ca_{12}Al_{14}O_{33}$ crystal allow to generation of broadband visible emission centered at 650 nm [29]. According to ionic space charge models, the oxygen molecules with a trapped electron are present on the surface of the Cr,Ca;YAG materials [27]. It is likely that the oxygens from space charge region were responsible for LIWE. This model also explains pressure dependence of LIWE. Probably, high ambient pressure suppresses electron transfer from the $Cr^{3+}/Cr^{4+}$ mixed valence pair to space charge region and, as result, bright visible emission was not observe.

More information on LIWE mechanism was extracted from measurement of power of transmitted laser beam through the Cr;YAG ceramic, hereinafter referred to as transmittance. The increase in transmittance during LIWE generation was caused by nonlinear scattering. The most interesting and important properties of nonlinear scattering are associated with the self-transparency phenomenon, which consists of light-induced optical homogenization of the initially turbid heterogeneous medium. Self-transparency phenomenon caused by a change of nonlinear refractive-index under the influence of a high intensity laser beam. The nonlinear refractive-index changes may be caused by different mechanisms such as the molecular-reorientation Kerr effect in liquids, the electronic Kerr effect in solids, thermal heating in both liquids and solids, and generation of nonequilibrium carriers in semiconductors and dielectrics [30].

We suppose that the generation of nonequilibrium carriers causes differences in the transmittance during displacement of the laser beam. This is consistent with literature since the LIWE phenomenon is accompanied by the generation and motion of free electrons [3,4,6,8]. The decrease in transparency due to the displacement of laser beam (fig. 8a) or an increase in ambient pressure (fig. 8b) can be explained by the phenomenon of self-transparency caused by generation of free electrons during LIWE. Based on this, the following conclusions can be drawn. First of all, an increase in ambient pressure suppresses the LIWE process as whole (electrons generation, radiative recombination et. all.), rather than quenching the radiative recombination process. Secondly, the low pressure and high density of the laser beam are insufficient to generate LIWE. In other words, multiphoton absorption and, as result, generation of free electrons does not occur when broadband visible emission is not observed.

*4.3 LIWE mechanism*

In summary, the following LIWE model is proposed. The Fig. 9 shows schematic illustration of mechanisms responsible for LIWE from transparent Cr;YAG ceramics. LIWE process take place in the following steps. The first stage is associated with the process of multiphoton absorption of infrared light by one of the Cr ions from a mixed valence pair (indicated by number 1). This leads to promoting of electron from $Cr^{3+}/Cr^{4+}$ mixed valence pair to space charge region (indicated by number 2). It should be noted that another valence state of chromium ions, as well as some structural defect, may be involved in this process. Moreover, the process of multiphoton absorption of infrared light by one of the Cr ions from a mixed valence pair occur, which leads to the separation of electron and promotion into conduction bands. This explains the difference in the number of photon involved in LIWE process, since various chromium ions, from mixed valence pair, can absorb infrared light. The interaction of this electron with the surface defects (oxygen from space charge region for example) leads to generation of broadband emission (indicated by number 3). And finally electron transfer to chromium ions from mixed valence pair (indicated by number 4). Several possible source of broadband emission can be proposed, such as electron-hole recombination [15] or oxygen from the space charge region [29].

Fig. 9 Schematic illustration of the mechanism responsible for LIWE from the transparent Cr:YAG ceramics.

It should be noted however that a more detailed explanation of the electron transfer and radiative recombination of the stored energy cannot be proposed. Therefore, the question about the LIWE mechanism remains open. Future work should therefore include the study of LIWE on transparent Yb;YAG ceramics.

**Acknowledgement.**

The authors are indebted to dr. M. Blees (CoorsTek, Netherlands) for making and providing the Cr:YAG samples, and also, for fruitful discussions. This work was supported by Polish National Science Centre, grant: PRELUDIUM-18 2019/35/N/ST3/01018.

## Conclusion

Strong laser induced white emission (LIWE) was observed from transparent Cr:YAG ceramics under vacuum upon 1064 nm excitation. The LIWE was detected only on the surface of the sample and wasn't in volume. The highest intensity of LIWE was obtained when the pressure in the measuring chamber was the lowest. An increase in ambient pressure above 0.5 mbar leads to a decrease in the intensity of LIWE. The emission spectra consisted of a wide band centred at 650 nm. It was also observed that the LIWE from transparent Cr:YAG ceramics was a threshold process exhibiting supralinear behavior. It was observed that the intensity of the LIWE increased nonlinearly with increasing incident laser power, with a strong increase in intensity after a certain power threshold that could be well fitted using a power law. The number of photons involved in broadband emission process was found in the range from 3 to 6. The decay time of LIWE was found to be 6±3 ms. The rise profiles demonstrate some instability after reaching the maximum intensity of broadband emission, which disappears after 8s. The rise time was found to be 9±6 ms.

In order to explain the nature of LIWE, the following two stages of this process have been proposed. The first stage is associated with the process of multiphoton absorption of infrared light by one of the Cr ions from a mixed valence pair. This leads to promoting of electron from $Cr^{3+}/Cr^{4+}$ mixed valence pair to space charge region. The interaction of this electron with the surface defects (oxygen from space charge region for example) leads to generation of broadband emission, and finally an electron transfer to chromium ions from mixed valence pairs. Several possible source of broadband emission can be proposed, such as electron-hole recombination [15] or oxygen from the space charge region [29].